\documentclass{article}

\usepackage[nonatbib,final]{neurips_2024}

\usepackage[utf8]{inputenc} 
\usepackage[T1]{fontenc}    
\usepackage{hyperref}       
\usepackage{url}            
\usepackage{booktabs}       
\usepackage{amsfonts}       
\usepackage{nicefrac}       
\usepackage{microtype}      
\usepackage{xcolor}         
\usepackage{graphicx}
\usepackage{wrapfig}
\usepackage{float}

\usepackage{amsmath}
\usepackage{amsfonts}
\usepackage{bm}

\def\eqref#1{equation~\ref{#1}}

\def\1{\bm{1}}

\DeclareMathAlphabet{\mathsfit}{\encodingdefault}{\sfdefault}{m}{sl}
\SetMathAlphabet{\mathsfit}{bold}{\encodingdefault}{\sfdefault}{bx}{n}

\title{Empathic Coupling of Homeostatic States\\ for Intrinsic Prosociality}

\author{%
  Naoto Yoshida \\ 
  Kyoto University\\
  \texttt{yoshida.naoto.8x@kyoto-u.ac.jp} \\
  \And
  Kingson Man \\
  Feeling Machines LLC \\
  \texttt{kingson@feelingmachines.net} \\
}

\begin{document}

\maketitle

\begin{abstract} 
When regarding the suffering of others, we often experience personal distress and feel compelled to help. Inspired by living systems, we investigate the emergence of prosocial behavior among autonomous agents that are motivated by homeostatic self-regulation. We perform multi-agent reinforcement learning, treating each agent as a vulnerable homeostat charged with maintaining its own well-being. We introduce an empathy-like mechanism to share homeostatic states between agents: an agent can either \emph{observe} their partner’s internal state (cognitive empathy) or the agent’s internal state can be \emph{directly coupled} to that of their partner’s (affective empathy). In three simple multi-agent environments, we show that prosocial behavior arises only under homeostatic coupling – when the distress of a partner can affect one’s own well-being. Our findings specify the type and role of empathy in artificial agents capable of prosocial behavior.
\end{abstract}


\section{Introduction}
 \vspace{-3mm}
 
For humans and other social animals, it is often distressing to regard the suffering of others. We feel empathy, sharing in the feelings of others rapidly and automatically through emotional contagion \cite{hsee1993emotional}. Such feelings can provide a strong motivation to reduce the suffering of others, even if it comes at some cost to the self. It has been proposed that tying one’s own welfare to the welfare of others can form the basis of prosocial behavior \cite{de2008putting}. 

\begin{wrapfigure}{r}{0.6\textwidth}
\vspace{-3mm}
  \centering
  \includegraphics[width=\linewidth,bb=0 0 929 324]{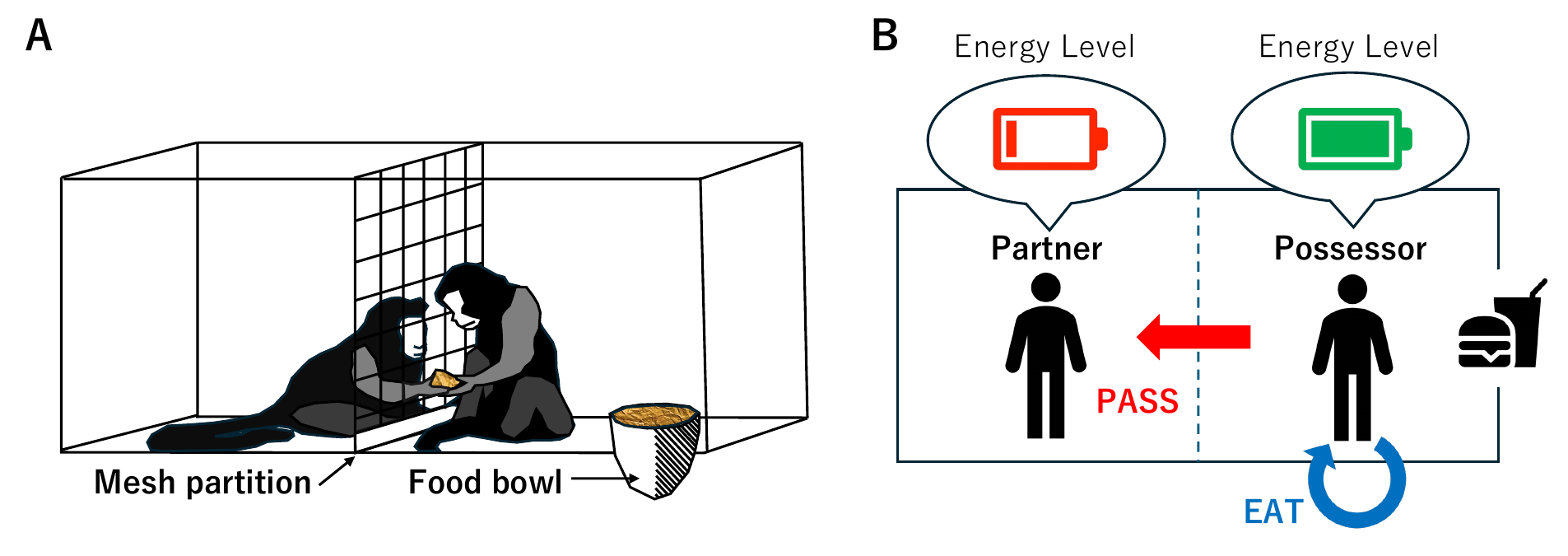}
  \vspace{-6mm}
  \caption{A: An experimental setup used in behavioral experiments to test the altruism of monkeys (illustration created based on \cite{de1997food}). B: A minimal reinforcement learning environment inspired by the behavioral experiment.}
  \label{fig:foodshare_env}
      \vspace{-5mm}
\end{wrapfigure}

Emotions and feelings, whether self- or other-directed, are theorized to arise from homeostasis, the regulation of internal body states within a range compatible with life \cite{damasio1999feeling}. Self-regulatory mechanisms have previously been implemented as a source of external motivation \cite{oudeyer2007typology,baldassarre2011intrinsic}. Here we regard homeostasis \cite{cannon1939wisdom} as an intrinsic and obligatory motivation of all living creatures. Homeostatic-like processes have recently been implemented in reinforcement learning (RL) agents and have resulted in the emergence of integrated behaviors \cite{yoshida2017homeostatic,yoshida2023homeostatic,yoshida2024emergence,yoshida2024synthesising}. There are many discussions of intrinsic motivation linked to artificial curiosity or exploration \cite{schmidhuber1991possibility,schmidhuber2010formal,baldassarre2011intrinsic,oudeyer2007typology,oudeyer2007intrinsic,der2012playful}. Also, homeostasis-based rewards have also been reported to facilitate exploration in the environment \cite{dulberg2023having}. In contrast to the discussion of the efficient exploration, here we start from the minimal condition for prosocial behavior originally proposed in the field of animal behavior. Under this condition, we show computationally that prosocial behavior does not arise from the individual's homeostatic reward alone. We then introduce the hypothetical homeostatic coupling term in the individual's reward function, which enables the emergence of the agent's prosocial behaviors. The necessity of this coupling term suggests that some form of intrinsic motivation beyond each individual's homeostasis is necessary for the emergence of prosocial behaviors in agents.

Here we extend these ideas to a model of social behavior using multi-agent RL, where each interacting agent is formulated as a {\bf homeostat} \cite{ashby1952design,seth2014cybernetic,man2019homeostasis}. We begin with the analysis of a multi-agent toy model inspired by behavioral experiments on monkeys to study prosocial behavior (Figure \ref{fig:foodshare_env}A), in which agents have the opportunity to share their own food resource with a needy conspecific (Figure \ref{fig:foodshare_env}B). We next propose some requirements for prosocial behavior. In several simulations, we compare the effects of different implementations of empathy \cite{christov2023preventing}, including cognitive empathy, in which an agent can observe the needy internal state of a partner, and affective empathy, in which an agent’s own internal states are coupled to their partner’s internal states. 

We contribute the following preliminary findings: 1) Even in a very simple system, prosocial behavior is not acquired when each agent aims only for homeostasis within its own body. 2) Prosocial behavior does not emerge even when an agent can directly observe the internal states of other agents (cognitive empathy). 3) Prosocial behavior was only observed when the agent’s internal state was directly coupled to the internal states of other agents (affective empathy). These results suggest that, for homeostats motivated by self-regulation, it is necessary to incorporate an additional homeostatic coupling parameter for prosocial behavior to arise. 

\subsection{Homeostatic Reinforcement Learning}
 \vspace{-3mm}

RL provides a framework to learn behavior in dynamic environments that maximizes the sum of future rewards emitted from the environment \cite{sutton2018reinforcement}. The objective of RL is to obtain a policy $\pi: {\cal S}\rightarrow {\cal A}$ that maximizes the expected value of the weighted cumulative sum of future rewards $\sum_{t=0}^\infty \gamma^t r_t$ for all states $s \in {\cal S}$, based on the experience of the interactions with an environment. Here, $t$ is the time step, $r$ is the reward signal, ${\cal S}$ is the set of states in the environment, ${\cal A}$ is the set of actions of the agent.  $0\leq \gamma < 1$ is a positive constant called the discount factor. 

Homeostatic RL \cite{keramati2011reinforcement,keramati2014homeostatic,hulme2019neurocomputational} integrates principles from physiological homeostasis by defining reward as the internally perceived reduction of deviations from homeostatic {\bf setpoints} \cite{keramati2017cocaine,juechems2019does,uchida2022computational,duriez2023homeostatic}. Concretely, the reward is defined as a quantity proportional to the temporal difference of the {\it drive} $D$, as $r_{t+1} = \beta (D_t - D_{t+1})$, where $\beta$ is the scaling constant \cite{keramati2011reinforcement}. The drive function $D (s^i)$ returns a value greater than or equal to zero, such as the distance between $s^i$ and $s^*$. Here $s^i$ is {\bf interoception} \cite{sherrington1906integrative}  that monitors the internal state of the agent's body \cite{sherrington1906integrative} and $s^*$ is the setpoint of the interoception. Homeostatic parameters are not arbitrarily defined, but are fundamental to the viability and functionality of the agent \cite{keramati2014homeostatic,man2019homeostasis}. In our conception, homeostatic RL asserts that the agent has a {\bf vulnerable} body. Vulnerability defined as the circular causality by which homeostatic states can affect the agent’s ability to regulate those states (i.e., it gets harder to take care of oneself as one falls apart). 

We used Proximal Policy Optimization (PPO, \cite{Schulmanetal_ICLR2016,schulman2017proximal}) as the RL optimizer in all of our experiments. The agent's policy model consisted of an encoder of inputs using a multi-layer perceptron, a recurrent connection using LSTM \cite{hochreiter1997long}, and softmax outputs for the categorical action probabilities in all experiments. Further experimental details are given in Appendix \ref{sec:hyper}.

\section{Experiment 1: Food Sharing Environment}
 \vspace{-3mm}

Inspired by ethology experiments on the altruism of monkeys (Figure \ref{fig:foodshare_env}A) \cite{de1997food}, we first created a minimal system for studying prosocial behavior. An overview of this environment is shown in Figure \ref{fig:foodshare_env}B. It has been reported that brown capuchins that are separated by a mesh will choose to share food when only one monkey has access to the food \cite{de1997food,de2000attitudinal}. This study explores the minimum configuration in which such sharing behavior occurs in autonomous agents.

In this environment, we assume two agents. The first is a passive agent called the `Partner', corresponding to the monkey in the left side of the cage with no direct access to food (Figure \ref{fig:foodshare_env}A). The other agent is the `Possessor', corresponding to the monkey on the right of Figure \ref{fig:foodshare_env}A, and who has access to food. Each agent has a binary energy state (High or Low). When the state is High, it transitions to Low with a small probability of $p=0.1$ at each time step. If the energy state is Low, it remains unchanged until the agent is able to eat food. If either one of the agents' energy states becomes Low and 10 steps have passed, the episode is failed and the environment is reset. In this environment, only the Possessor takes actions. The actions are EAT, causing the Possessor to transition to the High energy state, and PASS, causing the Possessor to transition to the High energy state. Further details of this environment are in Appendix \ref{sec:append_food}. 

A simple analysis of the environmental dynamics suggests that prosocial sharing behavior will not emerge if the Possessor is motivated only by its own homeostasis (Appendix \ref{sec:append_fooda}). Therefore, using this toy environment, we explore conditions under which the Possessor will share food with the Partner and prosocially maintain both agents’ energy states at High. We compare the following four conditions in numerical simulations: i) The Possessor optimizes only for its own homeostasis ({\bf none} condition). ii) The Possessor can observe the energy state of the Partner but is not specifically motivated to maintain the Partner's homeostasic state ({\bf cognitive} empathy condition). iii) The Possessor does not explicitly observe the energy state of the Partner, but has its own energy state coupled to the Partner’s energy state with a weighting factor ({\bf affective} empathy condition). Specifically, the weighting factor $w=0.5$ and the drive of the Possessor is given by $D = D_{\rm possessor} + w D_{\rm partner}$. iv) The final situation combines both cognitive and affective empathy. We used as in iii. The Possessor can explicitly observe the energy state of the Partner, and the Possessor’s energy state is coupled to that of the Partner ({\bf full} empathy condition).
\vspace{-3mm}

\subsection{Results}
 \vspace{-3mm}
 
\begin{figure}[t]
  \centering
  \includegraphics[width=0.8\linewidth]{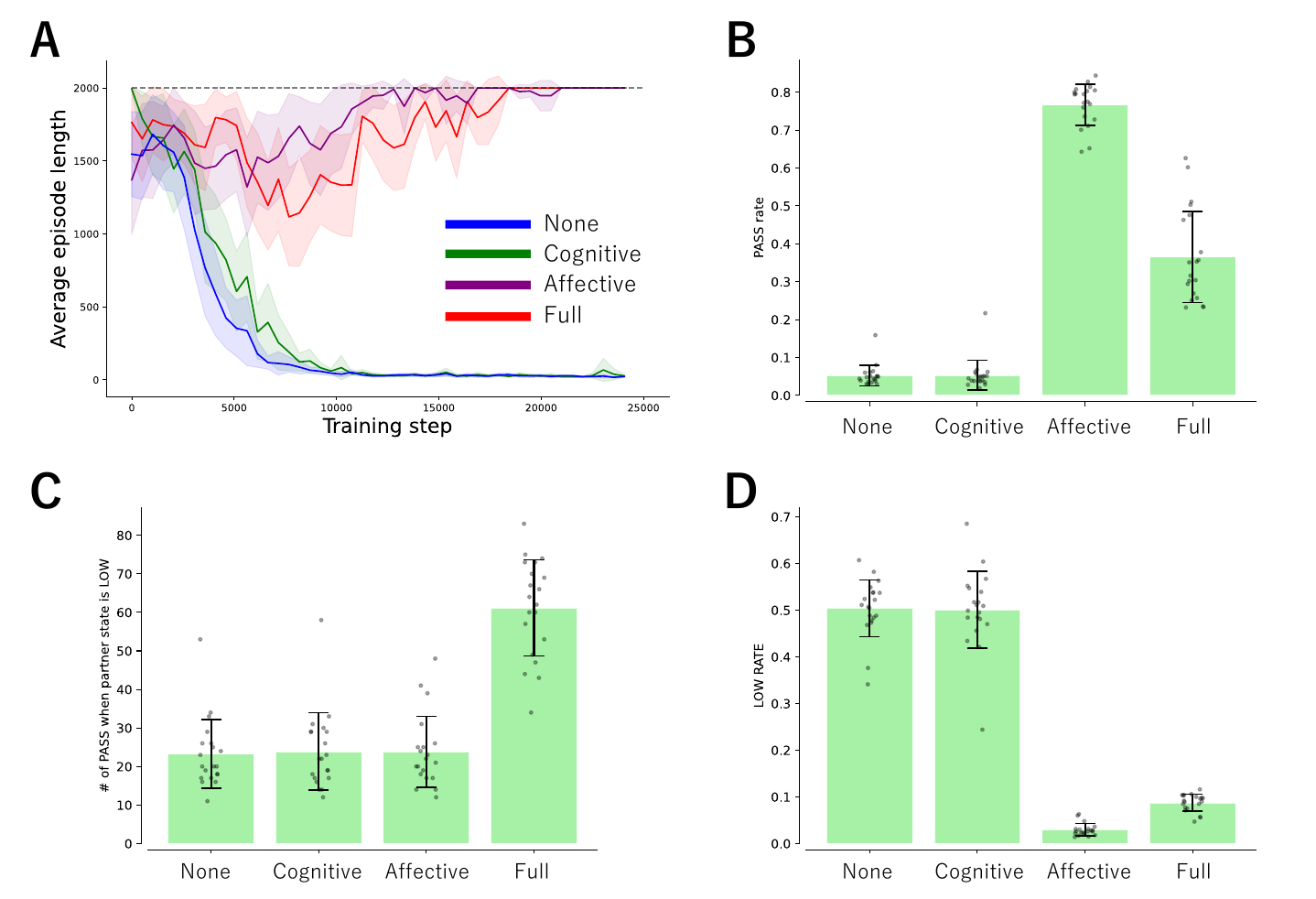}
  \caption{
Learning and behavior evaluation in a food sharing environment. A: Learning curves with performance measured by episode duration with both agents alive ({\emph n}=20, 95\% confidence intervals). Only the conditions that implement affective empathy (Affective and Full) result in long episode durations. B: PASS behavior selection rate out of 1,000 steps of the test run. Possessor agents in the Affective condition learn to frequently pass food to the Partner. C: Count of PASS actions when Partner is in the Low energy state. Possessor agents in the Full empathy condition learn to selectively pass food to the Partner when it is most needed. D: LOW state rate of Partner agent, out of 1,000 steps of the test run.
  }
  \label{fig:fs_results}
  \vspace{-3mm}
\end{figure}

Average learning curves are shown in Figure \ref{fig:fs_results}A. Performance is evaluated by episode duration, with a maximum length of 2000 steps. In the None condition, the PASS action is rarely selected \ref{fig:fs_results}B and episode lengths did not increase over training. Similar results are obtained in the Cognitive condition, in which the Possessor observes, but is not motivated by, the Partner's homeostatic state. On the other hand, the homeostatic states of both agents are maintained under the Affective and Full conditions, leading to long episode durations. In the Affective condition, the Possessor does not have explicit knowledge of the Partner’s energy state and so frequently chooses the PASS action to help the Partner maintain homeostasis, thereby also regulating its own homeostatic state because it is coupled to that of the Partner’s. This suggests that a strategy was acquired to maintain homeostasis between the two by supplying an excess of food to the Partner.

Figure \ref{fig:fs_results}C--D supports this speculation. Considering only the times when the Partner is in a Low energy state, Full condition agents selected the PASS action more often than Affective condition agents (Figure \ref{fig:fs_results}C). This implies that the Possessor in the full condition learned to select the PASS action only when the Partner's state was low. The values are the similar except for the full condition, but this is because there are few opportunities for the Partner's state to become LOW in the affective condition (Figure \ref{fig:foodshare_env}D). Altogether, these results suggest that a minimal requirement for prosocial behavior is an internalized motivation for the well-being of others.

\section{Experiment 2: Testing in Dynamic Environments}
\label{sec:continuous_env}
 \vspace{-3mm}

\begin{figure}[t]
  \centering
  \includegraphics[width=\linewidth,bb=0 0 1220 471]{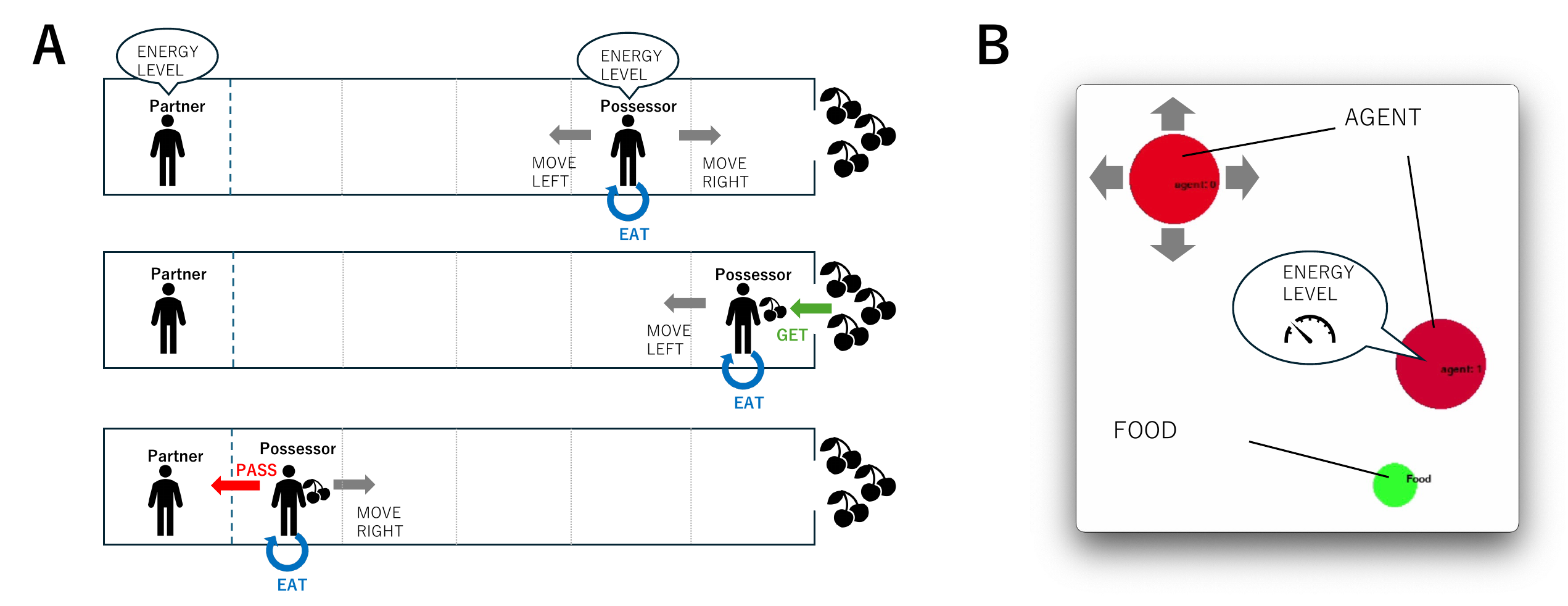}
  \caption{
Overview of the mobile agent environments. A: Linear grid environment. B: 2-D field environment. Detailed explanations are in Appendix \ref{sec:append_grid} and \ref{sec:append_2d}, respectively. 
  }
    \vspace{-3mm}
  \label{fig:grid_trap_env}
\end{figure}

Next, we investigated the generalizability of the findings from the food-sharing environment to 1-D and 2-D environments with mobile agents. The first is a linear grid environment (Figure \ref{fig:grid_trap_env}A), in which the Partner is, once again, trapped on the left side of the grid without access to food. The Possessor can acquire food at the far right side with the GET action, and increase their energy level with the EAT action. The Possessor can move LEFT and RIGHT to shuttle food to their Partner, and finally PASS it to the Partner when they are next to each other, increasing the Partner’s energy. Additionally, energy states are now represented as a continuous variable with a fixed rate of energy consumption. Further experimental details are in Appendix \ref{sec:append_grid}. 

The second mobile environment is one in which both agents can move on a two-dimensional field  (Figure \ref{fig:grid_trap_env}B). In this environment, there is no distinction between Partners and Possessors, as both agents can move and act freely. Food energy can be collected, consumed, and shared, as in the linear grid environment. However, if an agent's energy level decreases below some threshold, it becomes immobile and slowly starves. It then relies upon its Partner to share food in order to recover some energy and regain mobility. Further details, including on the small chance of random immobilization irrespective of energy level (‘injury’), are in Appendix \ref{sec:append_2d}.

\subsection{Results}
  \vspace{-3mm}

\begin{figure}[t]
  \centering
  \includegraphics[width=\linewidth,bb=0 0 1150 423]{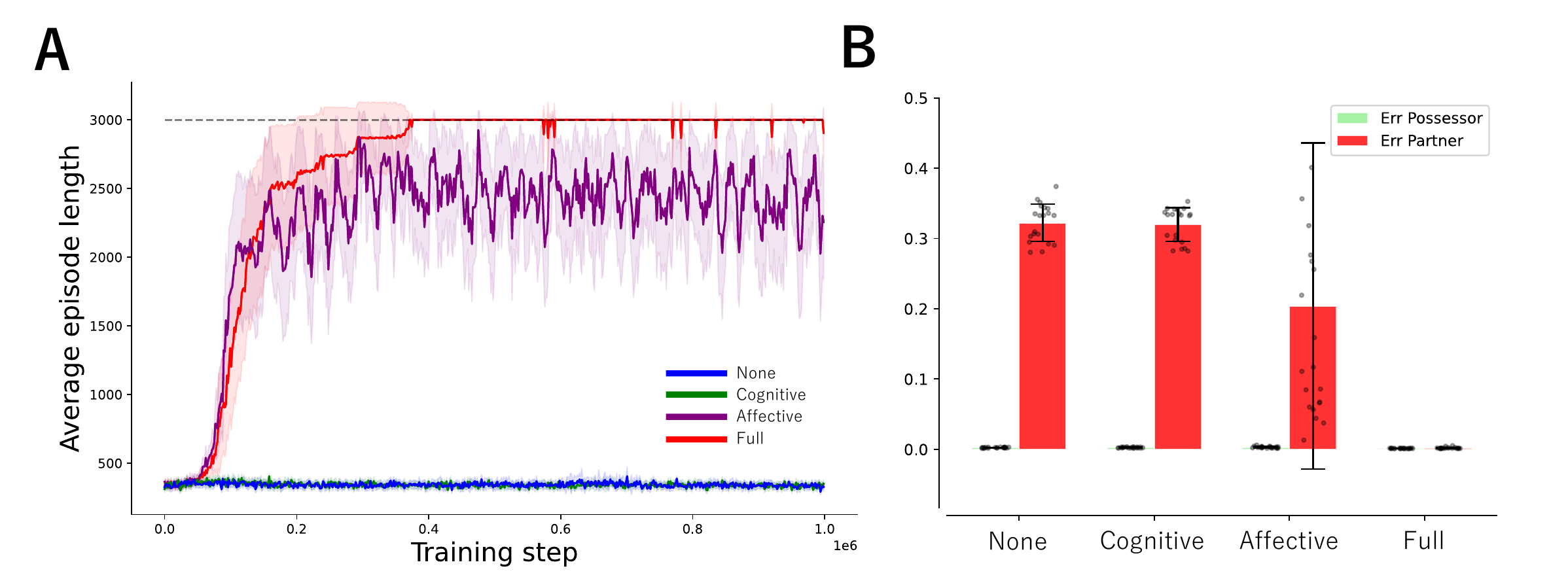}
      \vspace{-3mm}
  \caption{
Performance in the linear grid mobile environment. A: Learning curves with performance measured by episode duration with both agents alive ({\emph n}=20). B: Homeostatic drives of agents ($D_{\rm possessor}$ and $D_{\rm partner}$) averaged over 1000 timesteps.
  }
  \label{fig:grid_env}
\end{figure}

\begin{figure}[t]
  \centering
  \includegraphics[width=0.9\linewidth,bb=0 0 817 325]{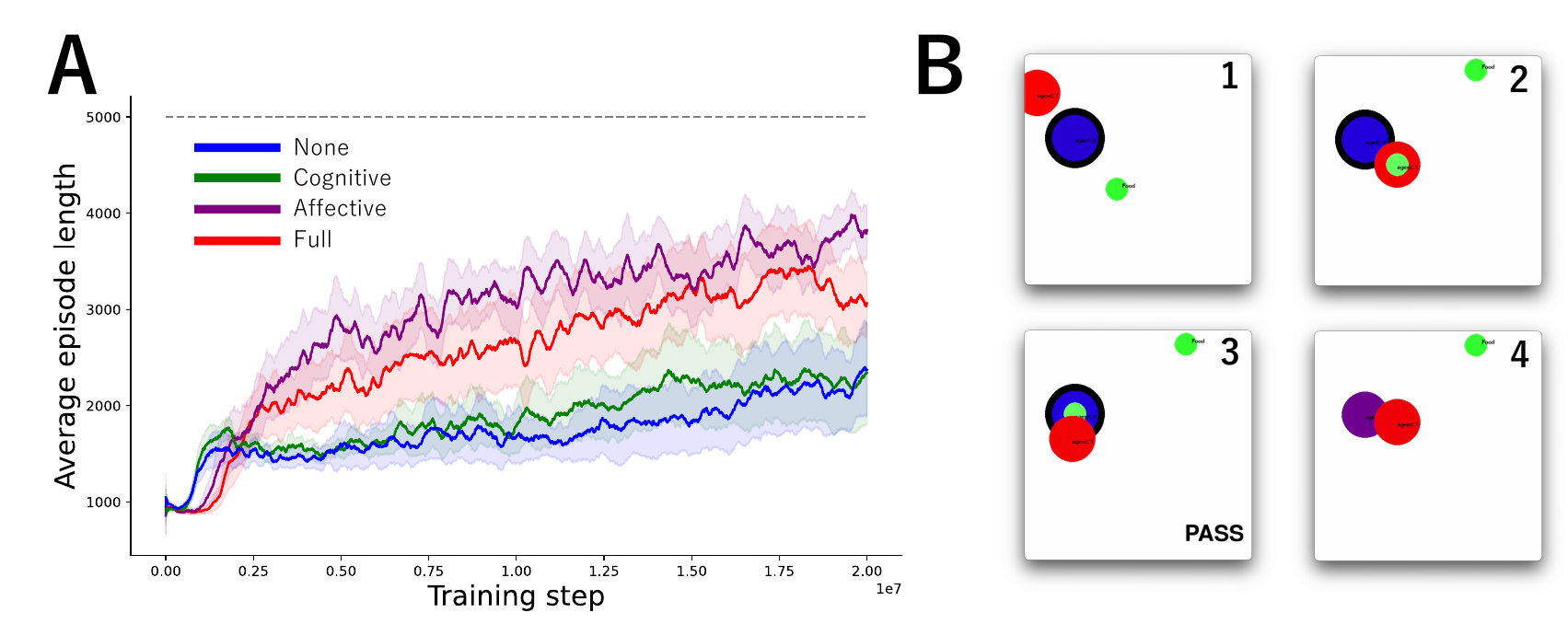}
      \vspace{-3mm}
  \caption{
Performance in the 2-D field mobile environment. A: Learning curves with performance measured by episode duration with both agents alive ({\emph n}=20). B: An example of helping behavior observed in the Affective condition. The action sequence progresses in order of the numbers in the top right corner of each panel.
  }
  \label{fig:trap_env}
\end{figure}

Figures \ref{fig:grid_env} and \ref{fig:trap_env} show the results of optimization in each mobile environment. No prosocial behavior was observed in the None and Cognitive conditions, and episode durations remained short (Figure \ref{fig:grid_env}A and \ref{fig:trap_env}A). As in Figure \ref{fig:grid_env}B, the variance of the homeostatic drive of the Partner ($D_{\rm partner}$) is large in the Affective condition. One possible explanation is that the Possessor agent cannot observe the energy state of its Partner, therefore the Partner agent is fed indiscriminately in the Affective condition, at various energy values (Appendix \ref{sec:appendix_hist}).  Figure \ref{fig:trap_env}B captures a sequence of prosocial behavior observed in the Affective condition. The blue agent is immobilized due to its low energy level. The red agent collects a green food pellet and returns to share it with the blue agent, turning it purple (replenishing some energy) and restoring it to mobility. 

\section{Discussions}
\vspace{-3mm}

This study investigated the emergence of prosocial behavior in simple RL agents motivated by homeostatic self-regulation. We found that prosocial behavior (food sharing) only occurred reliably under affective empathy, when the homeostatic states of agents were coupled. Perception of a partner’s state of need did not, on its own, drive prosocial behavior. The combination of cognitive and affective empathy in the Full condition drove more selective sharing behavior.  

Future research could explore more realistic empathy implementations, moving beyond giving agents direct access to partners' internal states. One possibility to achieve and maintain high group well-being is to implement a form of mutual information \cite{jaques2019social} in sequential social dilemmas \cite{leibo2017multi}, such that successfully self-regulating agents can influence each other and induce the well-being of others. Agents may also be designed to infer others' internal states from observable emotional behaviors. This process would better resemble the mirror neuron system, hypothesized to support emotion recognition and empathic behavior in humans and other animals \cite{rizzolatti2005mirror,iacoboni2009imitation}. For example, neurons in the inferior parietal lobule activate both during the observation and imitation of emotions; they can then trigger activity through the insula into the limbic system, known to activate during the firsthand experience of emotional feelings \cite{carr2003neural}. 

\section*{Acknowledgement}
\vspace{-3mm}
This work was supported by a grant from the Foresight Institute to KM, and Japan Society for the Promotion of Science KAKENHI grant 24K23892 to NY.

\newpage
\bibliographystyle{plain} 
\bibliography{reference}

\newpage
\appendix

\section{Agent Architectures in Experiments}
\label{sec:hyper}

\begin{wrapfigure}{r}{0.3\textwidth}
  \centering
  \includegraphics[width=\linewidth,bb=0 0 263 370]{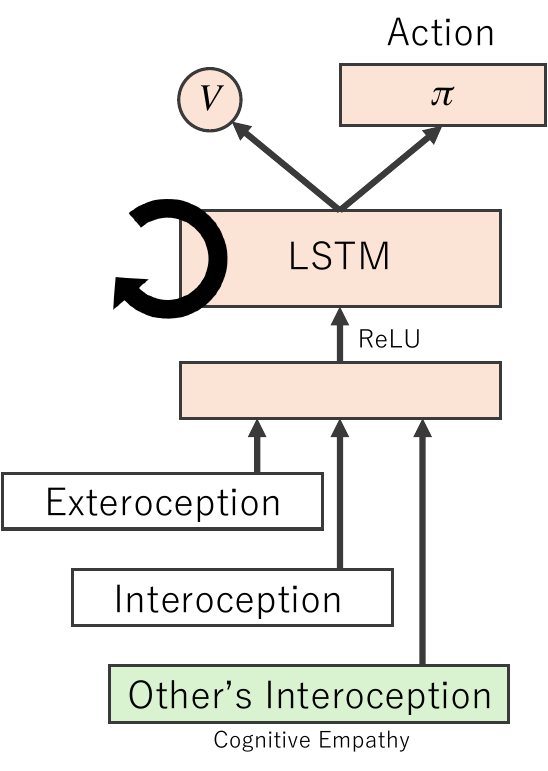}
  \caption{Basic network architecture of the agent in this study. Green observation is provided only when cognitive empathy is enabled ({\it cognitive} or {\it full} conditions).}
  \label{fig:model}
      \vspace{-5mm}
\end{wrapfigure}
In all of our computer experiments, we used Proximal Policy Optimization (PPO, \cite{Schulmanetal_ICLR2016,schulman2017proximal}) as the agent's optimizer. The agent's policy model consisted of an encoder of inputs using a multi-layer perceptron and a recurrent connection using LSTM \cite{hochreiter1997long} in all experiments (Figure \ref{fig:model}), and the action selection probability, $\pi$, was calculated by applying the softmax function to the affine transformation from the hidden state.
Value estimation $V_\pi$ is calculated as a 1D output by applying an affine transformation to the hidden state of the LSTM shared with the policy network. 

The models of the agents used in this study all had the same network architecture and optimization was performed with PPO.
Agents have Interoception for their own energy state as well as for observations from the outside world (Exteroception). In the {\it cognitive} condition, it also receives Interoception from other agents as well.
These observations are combined and input to the hidden layer with a linear mapping. The hidden layer takes the ReLU nonlinearity as the activation function and uses it as input to the LSTM.
A linear mapping from the output of the LSTM produces value predictions and categorical action selection probabilities using a softmax function. The table shows the network and PPO hyper--parameters for each experiment (Table \ref{tb:fs}--\ref{tb:2d}).

In the experiment in Section \ref{sec:continuous_env}, like previous study on multi-agent systems \cite{foerster2016learning}, we optimized the network weights and shared the acquired experience with all agents to facilitate learning.

\begin{table}[H]
\caption{Hyper-parameters of Food Sharing Environment}
\begin{center}
\begin{tabular}{c|c}
Exteroception dim & none  \\
Interoception dim & 1 (energy state)  \\
Interoception dim of other agent (cognitive empathy) & 1 (energy state)  \\
Hidden dim &  16\\
LSTM hidden state dim  &  16 \\
Total time steps & 25,000\\
Learning rate & 0.001 \\
Number of parallel sampling threads & 16\\
Sampling steps & 32\\
Discount factor ($\gamma$)& 0.99 \\
GAE lambda & 0.95\\
Number of minibatches & 4\\
Update epochs & 4\\
Normalizing advantage & True\\
Clip coefficient of policy update & 0.1 \\
Value clipping loss & True \\
Entropy coefficient & 0.01\\
Value loss coefficient & 0.5\\
Maximum gradient norm & 0.5\\
\end{tabular}
\end{center}
\label{tb:fs}
\end{table}

\begin{table}[H]
\caption{Hyper-parameters of Grid Environment}
\begin{center}
\begin{tabular}{c|c}
Exteroception dim & 5 (position, one-hot) + 2 (having food flag, one-hot)  \\
Interoception dim & 1 (energy state)  \\
Interoception dim of other agent (cognitive empathy) & 1 (energy state)  \\
Hidden dim&  32\\
LSTM hidden state dim  & 32 \\
Total time steps & 1,000,000\\
Learning rate & 0.001 \\
Number of parallel sampling threads & 16\\
Sampling steps & 100\\
Discount factor ($\gamma$)& 0.99 \\
GAE lambda & 0.95\\
Number of minibatches & 4\\
Update epochs & 4\\
Normalizing advantage & True\\
Clip coefficient of policy update & 0.1 \\
Value clipping loss & True \\
Entropy coefficient & 0.01\\
Value loss coefficient & 0.5\\
Maximum gradient norm & 0.5\\
\end{tabular}
\end{center}
\label{tb:grid}
\end{table}

\begin{table}[H]
\caption{Hyper-parameters of 2D Field Environment}
\begin{center}
\begin{tabular}{c|c}
Exteroception dim & 2 (position) + 2 (food position) \\
& + 2 (having food flag, one-hot) + 2 (movable flag, one-hot)\\
Interoception dim & 1 (energy state)  \\
Interoception dim of other agent (cognitive empathy) & 1 (energy state)  \\
Hidden dim&  64\\
LSTM hidden state dim  &  64 \\
Total time steps & 20,000,000\\
Learning rate & 0.001 \\
Number of parallel sampling threads & 16\\
Sampling steps & 1024\\
Discount factor ($\gamma$)& 0.99 \\
GAE lambda & 0.95\\
Number of minibatches & 2\\
Update epochs & 4\\
Normalizing advantage & True\\
Clip coefficient of policy update & 0.1 \\
Value clipping loss & True \\
Entropy coefficient & 0.0\\
Value loss coefficient & 0.3\\
Maximum gradient norm & 0.5\\
\end{tabular}
\end{center}
\label{tb:2d}
\end{table}

\newpage

\section{Details of Environments}
\subsection{Food Sharing Environment}
\label{sec:append_food}
The Possessor has two actions. One is EAT, and the Possessor can recover the energy state of the agent described below by eating food. The other action is PASS, and the Possessor can feed the Partner. Both of these agents have a binary energy state (High and Low), and when the state is High, it changes to Low at a small probability of $p=0.1$ at each time step. If the energy state is Low, it is left unchanged at each time step, and only changes to High if the agent is able to eat the food. At the start of the environment, both energy states are randomly determined, and if either one of the agents' energy states becomes Low and 10 steps have passed, the environment is reset for both agent.

In this environment, only the action optimization of the Possessor is possible. The drive for the homeostasis of the Possessor is $D_{\rm possessor}=-\ln P^*(s^i_t)$. Here, $s^i_t \in \{{\rm High}, {\rm Low} \}$ represents the agent's interoception at time $t$, and $P^*(\cdot)$ is a probability distribution representing the desirability of each state, with $P^*({\rm High})=0.95$ and $P^*({\rm Low})=0.05$. Therefore, the Possessor aims for homeostasis, preferring $s^i={\rm High}$ over $s^i={\rm Low}$.

\subsection{Grid Environment}
\label{sec:append_grid}
The first is a grid environment (Figure \ref{fig:grid_trap_env}A), in which the Partner is fixed to the left side of the grid, just like in the Food Sharing environment. The Possessor can access the food area on the right side. The Possessor can move left and right and know their position (there are five positions in the environment). The Possessor can acquire food by arriving at the right side and selecting the GET action. As a result, the Possessor has two states: with food and without food.
At this point, the Possessor can also eat the food (by selecting the EAT action), or move while holding the food and PASS it to the Partner. The Possessor can choose between five actions: move left or right, EAT, GET, and PASS.

In addition, the energy state of each agent is represented as a continuous variable. The dynamics of the energy state $s^i$ is represented by $s^i_{t+1} = s^i_t - \delta_d + I_t$. In this case, $\delta_d = 0.003$ is a fixed constant that represents a certain amount of energy consumption. $I_t$ is a function that returns 0.1 when the agent has ingested food, and 0 otherwise. In this experimental system, the drive function was given by the squared error $D=\|s^i \|^2$, and the agents were trained with a learning rate of $\beta=100$. If the energy state of any of the agents deviated from the range $[-1, 1]$, the episode was terminated.

\subsection{2D Field Environment}
\label{sec:append_2d}
This environment is one in which the agents can move around in a two-dimensional continuous space (Figure \ref{fig:grid_trap_env}B). In this environment, there is no distinction between Partners and Possessors, and both agents can move around and consume food.  The energy state changes in the same way as in the grid environment (with a dynamics of $\delta_d = 0.001$, and energy is restored by 0.3 when food is consumed). 

In addition, each agent can carry food, and if it is close enough to another agent while carrying food, it can give the food to the other agent. Also, in this environment, if an agent's energy level becomes less than -0.7, it is considered to be damaged and cannot move. Therefore, in such a situation, the agent needs to be helped to be able to move again by having another agent bring it food. Furthermore, when both agents are able to move, each agent encounters an accident at a small probability $p=0.0005$ at each time. If an agent encounters an accident, its energy value immediately becomes -$0.7$, and the agent becomes immobile. 

The scaling of the drive function and reward was the same as in the grid environment. In training in this environment, all agents were trained in a situation where the network weights were shared.

\newpage
\section{State-Transition Diagram of Food Sharing Environment}
\label{sec:append_fooda}
\begin{figure}[H]
  \centering
  \includegraphics[width=0.7\linewidth,bb=0 0 624 512]{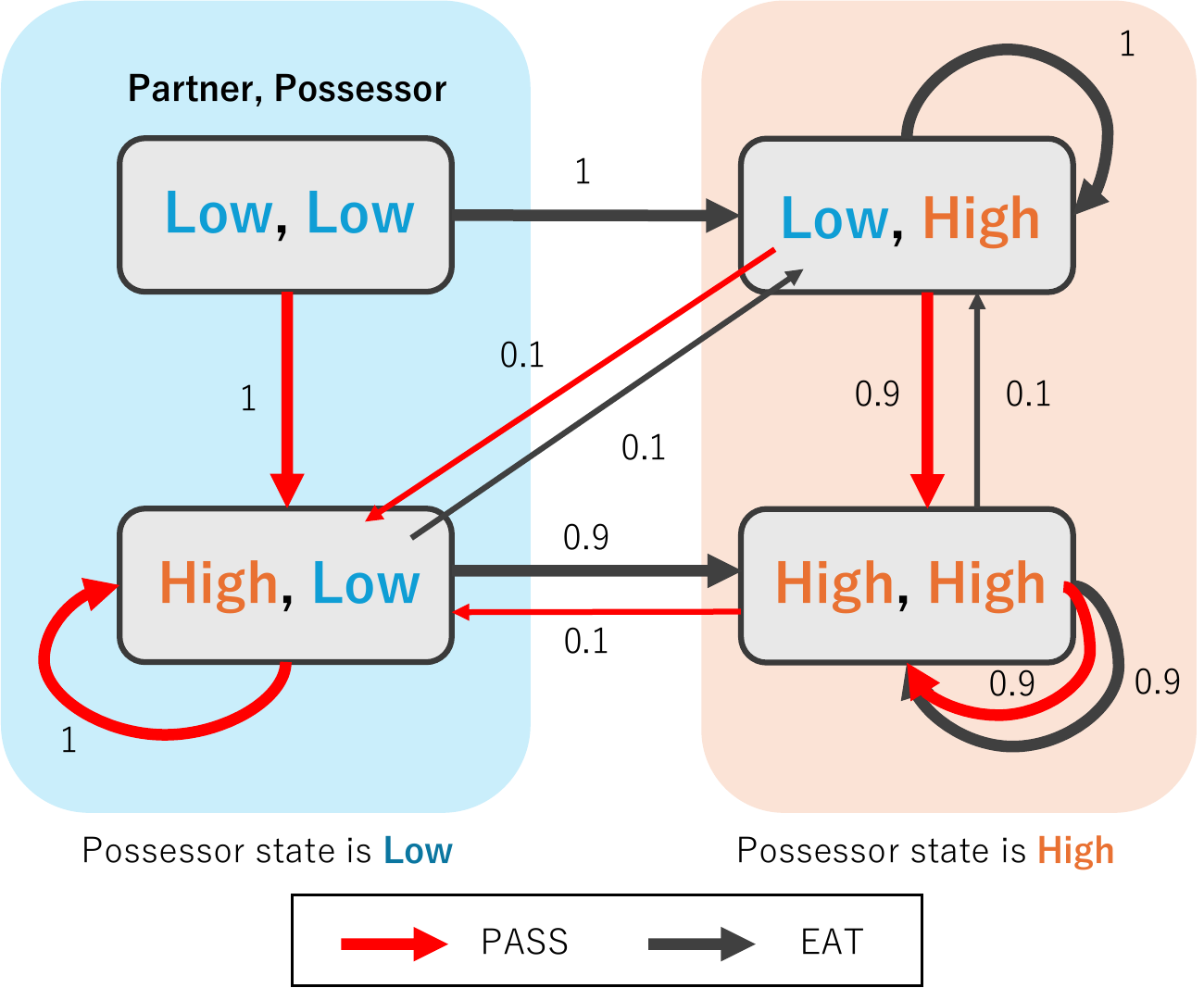}
  \caption{
  State transition diagram of the food sharing environment.
  }
  \label{fig:afs}
      \vspace{-5mm}
\end{figure}

All the state transitions in the Food Sharing environment (Figure \ref{fig:afs}). From this figure, we can see that there is always a risk of the internal state transitioning from High to Low when the Possessor chooses the PASS action. This can be seen from the transitions of the blue and red macro states (corresponding to the Possessor's interoception) on the left and right when the PASS action is chosen. Therefore, it is always optimal for the homeostasis of the Possessor alone to choose the EAT action, and it is suggested that in such a situation, no action to help the Partner will emerge.
 
\newpage
\section{Histogram showing the difference between the {\it affective} and {\it full} conditions in the Grid environment}
\label{sec:appendix_hist}

\begin{figure}[h]
  \centering
  \includegraphics[width=\linewidth,bb=0 0 659 479]{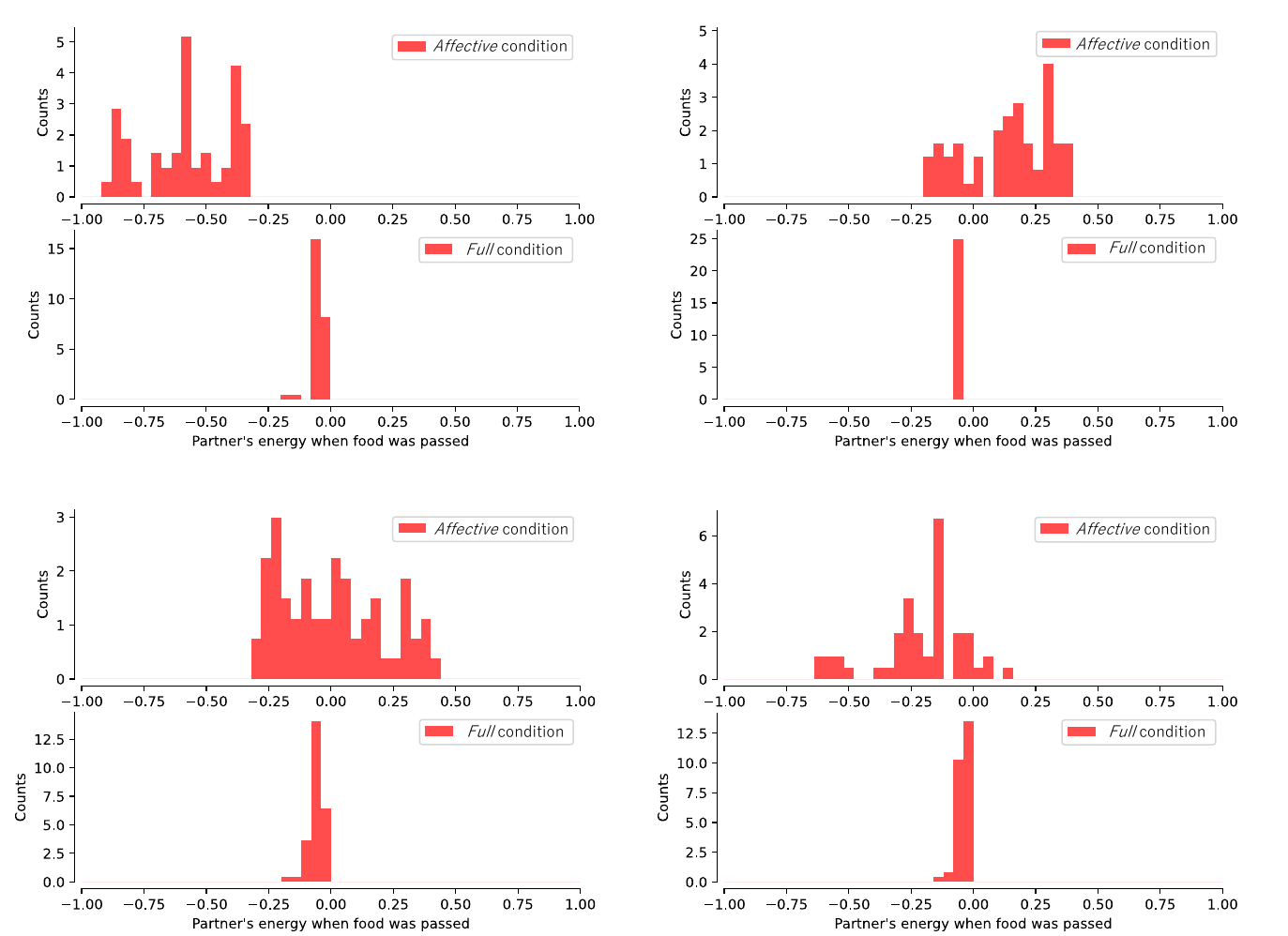}
  \caption{
  Typical histograms of the energy state of the Partner agent when it ingests food during the 2000-step test run after optimization. 
  }
\end{figure}


\end{document}